\documentclass[aps,prd,amsmath,twocolumn,superscriptaddress,preprintnumbers,amssymb,showpacs,floatfix,nofootinbib,longbibliography]{revtex4-1}
\pdfoutput=1

\usepackage{graphicx}
\usepackage{bm}
\usepackage{hyperref}
\usepackage{hyperref}
\usepackage{slashed}
\usepackage{color}
\usepackage{aas_macros}
\usepackage{slashed}
\usepackage{lipsum}
\usepackage{subfigure}
\usepackage{multirow}
\usepackage{amsmath}
\usepackage{array} 
\usepackage{varwidth} 

\newcommand{\be}{\begin{equation}}
\newcommand{\ee}{\end{equation}}
\newcommand{\bea}{\begin{eqnarray}}
\newcommand{\eea}{\end{eqnarray}}

\hypersetup{
     colorlinks   = true,
     citecolor    = magenta,
     urlcolor     = magenta,
     linkcolor    = magenta
}

\usepackage{listings}
\usepackage{color,xcolor}

\begin{document}

\preprint{SLAC-PUB-17586}

\title{Celestial-Body Focused Dark Matter Annihilation Throughout the Galaxy}

\author{Rebecca K. Leane}
\thanks{{\scriptsize Email}: \href{mailto:rleane@slac.stanford.edu}{rleane@slac.stanford.edu}; {\scriptsize ORCID}: \href{http://orcid.org/0000-0002-1287-8780}{0000-0002-1287-8780}}
\affiliation{SLAC National Accelerator Laboratory, Stanford University, Stanford, CA 94039, USA}

\author{Tim Linden}
\thanks{{\scriptsize Email}: \href{mailto:linden@fysik.su.se}{linden@fysik.su.se}; {\scriptsize ORCID}: \href{http://orcid.org/0000-0001-9888-0971}{0000-0001-9888-0971}}
\affiliation{Stockholm University and The Oskar Klein Centre for Cosmoparticle Physics,  Alba Nova, 10691 Stockholm, Sweden}

\author{Payel Mukhopadhyay}
\thanks{{\scriptsize Email}: \href{mailto:payelmuk@stanford.edu}{payelmuk@stanford.edu}; {\scriptsize ORCID}: \href{http://orcid.org/0000-0002-3954-2005}{0000-0002-3954-2005}}
\affiliation{SLAC National Accelerator Laboratory, Stanford University, Stanford, CA 94039, USA}
\affiliation{Physics Department, Stanford University, Stanford, CA 94305, USA}

\author{Natalia Toro}
\thanks{{\scriptsize Email}: \href{mailto:ntoro@slac.stanford.edu}{ntoro@slac.stanford.edu}; {\scriptsize ORCID}: \href{http://orcid.org/0000-0002-8150-3990}{0000-0002-8150-3990}}
\affiliation{SLAC National Accelerator Laboratory, Stanford University, Stanford, CA 94039, USA}

\begin{abstract}
Indirect detection experiments typically measure the flux of annihilating dark matter (DM) particles propagating freely through galactic halos. We consider a new scenario where celestial bodies ``focus" DM annihilation events, increasing the efficiency of halo annihilation. In this setup, DM is first captured by celestial bodies, such as neutron stars or brown dwarfs, and then annihilates within them. If DM annihilates to sufficiently long-lived particles, they can escape and subsequently decay into detectable radiation. This produces a distinctive annihilation morphology, which scales as the product of the DM and celestial body densities, rather than as DM density squared. We show that this signal can dominate over the halo annihilation rate in $\gamma$-ray observations in both the Milky Way Galactic center and globular clusters. We use \textit{Fermi} and H.E.S.S. data to constrain the DM-nucleon scattering cross section, setting powerful new limits down to $\sim10^{-39}~$cm$^2$ for sub-GeV DM using brown dwarfs, which is up to nine orders of magnitude stronger than existing limits. We demonstrate that neutron stars can set limits for TeV-scale DM down to about $10^{-47}~$cm$^2$.
\end{abstract}

\maketitle

\section{Introduction}

Celestial bodies provide versatile environments to discover new physics. Peppered throughout the Galaxy, their large abundances can be used to collectively power a bright dark matter (DM) annihilation signal.

Previous studies have examined DM particles that scatter in celestial bodies and become gravitationally bound.  The trapped DM can heat the objects that capture it, with contributions from both the DM kinetic energy and the absorption of DM annihilation products by the capturing body. The latter, dominant source of heating relies on the DM annihilation products either interacting or decaying within the celestial body.  For neutron stars (NS), this DM heating signal has been studied in e.g. Refs.~\cite{Goldman:1989nd,
Kouvaris:2007ay,
Bertone:2007ae,
deLavallaz:2010wp,
Kouvaris:2010vv,
Guver:2012ba,
McCullough:2010ai,
Baryakhtar:2017dbj,
Raj:2017wrv,
Bell:2018pkk,
Garani:2018kkd,
Chen:2018ohx,
Dasgupta:2019juq,
Hamaguchi:2019oev,
Camargo:2019wou,
Bell:2019pyc,
Garani:2019fpa,
Acevedo:2019agu,
Joglekar:2019vzy,
Joglekar:2020liw,
Bell:2020jou,
Dasgupta:2020dik,
Garani:2020wge}. DM heating using the full BD population was considered recently in Ref.~\cite{Leane:2020wob}. NSs and BDs are both efficient accumulators of DM, due to being relatively dense, and in the case of BDs, very large.  

A complementary approach arises when DM annihilates to long-lived mediators. In this scenario, the mediator can escape the celestial body and  decay to observable final states. Long-lived particles appear naturally in many well-motivated extensions of the SM ~\cite{Kobzarev:1966qya,Okun:1982xi,Holdom:1985ag,Holdom:1986eq,Batell:2009zp,Martin:1997ns}, with an extensive search program~\cite{Reece:2009un,Morrissey:2014yma,Aad:2015rba}. Dark sectors with long-lived mediators have previously been considered in local searches of the Sun~\cite{Batell:2009zp,Pospelov:2007mp,Pospelov:2008jd,Rothstein:2009pm,Chen:2009ab,Schuster:2009au,Schuster:2009fc,Bell_2011,Feng:2015hja,Kouvaris:2010,Feng:2016ijc,Allahverdi:2016fvl,Leane:2017vag,Arina:2017sng,Albert:2018jwh, Albert:2018vcq,Nisa:2019mpb,Niblaeus:2019gjk,Cuoco:2019mlb,Serini:2020yhb,Mazziotta:2020foa} and Earth~\cite{Feng:2015hja}.

We consider, for the first time, DM annihilation to long-lived particles in NSs and BDs, which are advantageous systems due to their superior and complementary scattering cross-section sensitivity. This allows us to consider a new type of annihilation signal: one ``focused" by the population of celestial bodies. DM is first efficiently captured in dense NSs or BDs. As the DM density increases, DM annihilation inside the object becomes efficient. The DM annihilation proceeds through a long-lived particle which escapes the celestial body and subsequently decays, producing a flux detectable at Earth.

Our signal exploits the fact that celestial bodies exist in large quantities in the inner Galaxy~\cite{Freitag:2006qf,Generozov:2018niv,Kim:2018llc}, as well as other DM-dense environments such as globular clusters~\cite{Pfahl:2001df,Ivanova:2008,Ye:2019luh}. Notably, while DM annihilation in the halo scales as the DM density squared, the celestial-body focused annihilation rate scales as a single power in DM number density (assuming equilibration between the annihilation and capture rates in a given object) multiplied by the celestial-body number density.  This distinctive scaling can potentially disentangle the origin of an observed DM annihilation signal. Moreover, because the DM density within the celestial body can become extremely high, our scenario potentially provides a more sensitive probe than halo annihilation, especially for large DM masses or suppressed annihilation cross sections, such as $p-$wave annihilation~\cite{Slatyer:2017sev,Leane:2020liq}.

In this paper, we investigate the relative strength of celestial-body-focused annihilation compared to DM annihilation in the Milky Way halo. We compare our results with existing $\gamma$-ray data, and produce new limits on DM annihilation to long-lived particles. We identify two environments where a NS-focused or BD-focused annihilation signal can dominate over halo annihilation. These are the Galactic center, which is extremely DM dense, and globular clusters, which can have not only large DM densities, but also low DM velocity dispersions, allowing more DM to be captured by the celestial body.

This paper is organized as follows. In Section~\ref{sec:setup}, we review DM capture and annihilation in celestial bodies, and detail the long-lived mediator model. We then discuss the Milky Way Galactic center signal in Section~\ref{sec:mw}, and the resulting constraints in Section~\ref{sec:results}. We discuss the globular cluster signal in Section~\ref{sec:globular}. We discuss the implications of our results in Section~\ref{sec:conclusion}.

\section{Setup}
\label{sec:setup}
\subsection{Dark Matter Capture in Celestial Bodies}
\label{sec:capture}

As in the case of standard halo annihilation, the strength of the NS- and BD-focused signal depends on the DM density in the object's environment. However, for BD/NS-focused annihilation, the DM density directly determines the rate of DM capture onto celestial bodies, an interaction that is only linearly (rather than quadratically) dependent on the DM density. Here, we use a generalized Navarro-Frenk-White (NFW) density profile, which is defined as a function of galactic radius $r$,~\cite{Navarro:1996}
\begin{equation}
 \rho_\chi(r)=\frac{\rho_0}{(r/r_s)^\gamma(1+(r/r_s))^{3-\gamma}},
 \label{eq:DM_density}
\end{equation}
where $r_s$ is the scale radius, $\rho_0$ is normalized to the local DM density value, and $\gamma$ is the index that determines the inner slope of the DM profile.

DM from the Galactic halo can fall onto a celestial object, encountering the surface after being sped up to approximately the object's escape velocity, \mbox{$v_{\rm esc} = \sqrt{{2G_N M}/{R}}$}, where $G_N$ is the gravitational constant, $M$ is the mass of the object, and $R$ is the object's radius.
As the DM particle transits through the object, it can scatter with the stellar material and lose energy. Once the kinetic energy of the DM is less than the gravitational potential, the DM particle is captured. DM capture can occur via single or multiple scatters with the stellar constituents \cite{Kouvaris:2010,Bramante:2017,Dasgupta:2019,Ilie:2020}.

The largest possible rate of capture is obtained by assuming that  all DM that passes through the effective area of the BD/NS is captured. This ``maximum capture rate'' (sometimes also referred to as ``geometric capture rate'') is given by~\cite{Garani:2017jcj}
\begin{equation}
\label{eq:max_capture}
    C_{\rm max} = \pi R^2 n_\chi (r) v_0 \left(1+\frac{3}{2}\frac{v_{\rm esc}^2}{\overline{v}(r)^2} \right) \xi(v_p,\overline{v}(r)),
\end{equation}
where $\bar{v}$ is the DM velocity dispersion, $n_\chi(r)$ is the DM number density profile, related to Eq.~\ref{eq:DM_density} via \mbox{$n_\chi(r)=\rho(r)/m_\chi$, $v_0=\sqrt{8/(3\pi)}\overline{v}$}, and $\xi(v_p,\overline{v}(r))$ takes into account the motion of the compact object with respect to the DM halo (this is $\sim1$ for our scenarios, and we neglect it in what follows).

In practice, a BD/NS is not perfectly efficient at capturing DM: some DM particles impinging on the star will not scatter, and others will have sufficient energy after scattering that they are not captured.  The capture rate therefore depends on both the DM scattering cross section off the constituents of the celestial body and the kinematics of these scatters.  A treatment that accounts for  both single and multiple scatters in DM capture was developed in Ref.~\cite{Bramante:2017}.

The probability for a given DM particle to undergo $N$ scatters is
\begin{equation}
    p_N(\tau) = 2 \int_0^1 dy \frac{y e^{-y \tau} (y \tau)^N}{N!},
\end{equation}
where $\tau$ is the optical depth,
\begin{equation}
    \tau=\frac{3}{2}\frac{\sigma}{\sigma_{\rm sat}}.
\end{equation}
Here $\sigma_{\rm sat}$ is the saturation cross section of DM capture onto nucleons, and is given by $\sigma_{\rm sat} = {\pi R^2}/{N_n}$ where $N_n$ is the number of nucleons. 

The total capture rate in this formalism for a single celestial body is then given by
\begin{equation}
    C = \sum_{N= 1}^{\infty} C_N ,
\label{eq:multiscatter_total}
\end{equation}
where $C_N$, defined below, is the capture rate associated with particles that scatter $N$ times.  In practice, this sum can be truncated at a maximum finite $N\gg\tau$, because scattering more than $\tau$ times is exponentially suppressed. 
The rate for a particle to impinge on the body, scatter $N$ times, and lose enough energy in the process to become trapped in the star is given by\footnote{For NSs, blue-shifted incoming DM velocities are included by replacing $v_{\rm esc}\rightarrow\sqrt{2\chi}$, where \mbox{$\chi=1-\sqrt{1-2G_N M/R}$}.}
\begin{align}
\label{eq:multi scatter}
C_N &= \frac{\pi R^2 p_N(\tau)}{(1-2G_N M/R)} \frac{\sqrt{6} n_{\chi}}{3 \sqrt{\pi} \bar{v}} \times\\ &\left((2 \bar{v}^2 + 3 v_{\rm esc}^2) - (2 \bar{v}^2 + 3 v_{N}^2)\exp \left(-\frac{3(v_N^2 - v_{\rm esc}^2)}{2 \bar{v}^2}\right) \right)\nonumber,
\end{align}
where the term $v_N = v_{\rm esc} (1 - \beta_{+}/2)^{-N/2}$ with $\beta_{+} = {4 m_{\chi} m_n}/{(m_{\chi} + m_n)^2}$ takes into account energy lost by DM in each scattering event.  For sufficiently large $N$, $v_N^2 - v_{\rm esc}^2$ becomes much larger than $\bar{v}^2$ and $C_N$ in Eq.~\ref{eq:multi scatter} rapidly approaches $p_N \times C_{\rm max}$ (neglecting the $\xi$ factor in Eq.~\ref{eq:max_capture}). In other words, particles that undergo $N$ scatters are efficiently captured.  As $\tau$ increases above this minimum number of scatters required for efficient capture, the capture rate $C$ in Eq.~\ref{eq:multiscatter_total} asymptotes to the maximum capture rate given by Eq.~\ref{eq:max_capture}. 

We note that DM capture will be truncated for sufficiently light DM masses, because DM can rapidly evaporate out from the system. Evaporation occurs if the core of the system has both sufficiently high temperatures (which impart kinetic energy to the DM), and sufficiently low gravitational potential. From Ref.~\cite{Leane:2020wob}, we expect an evaporation mass of around a few MeV for BDs. For NSs, the DM evaporation mass is $\sim300~$eV for old NSs which have cooler cores~\cite{Bell:2020lmm}, and we estimate up to $\sim$MeV for very young NSs. As we will (arbitrarily) consider DM masses above 10 MeV, the evaporation mass will be lighter than our range of interest.

To calculate the total expected capture rate from the full celestial body population in a given system (e.g. the Galactic center or globular clusters), we also need to take into account the number density of the object in the region of interest, $n_{\rm BD/NS}$. In this scenario, the total DM capture rate by a population of BDs/NSs, $C_{\rm BD/NS, tot}$, can be written as 
 \begin{equation}
     C_{\rm BD/NS, tot} = 4 \pi \int ^{r_2}_{r_1} r^2 \, n_{\rm BD/NS} \, C \,  dr,
\label{eq:tot_capture}
\end{equation}
where $C$ is the capture rate by a single BD or NS, $n_{\rm BD/NS}$ is the BD or NS number density, and $r$ is the radial distance from the center of the system. This total capture rate of the full population of celestial bodies will be computed in Sec. \ref{sec:mw}, \ref{sec:results} and \ref{sec:globular}. In the following subsection, we compare cross sections for DM capture in a \textit{single} celestial body.

\subsection{Comparing Different Celestial Targets}

To determine which type of celestial body will dominate the results for a given mass or cross-section sensitivity range, we compare choices of different objects. The optimal target can be chosen based on the system's core temperatures (lower core temperatures provide less kinetic energy for DM to escape, potentially providing more sensitivity to lower DM masses), and the system's density (which increases the probability of DM capture for small DM/nucleon cross sections).

\begin{figure} [t!]
    \centering
    \includegraphics[width=\columnwidth]{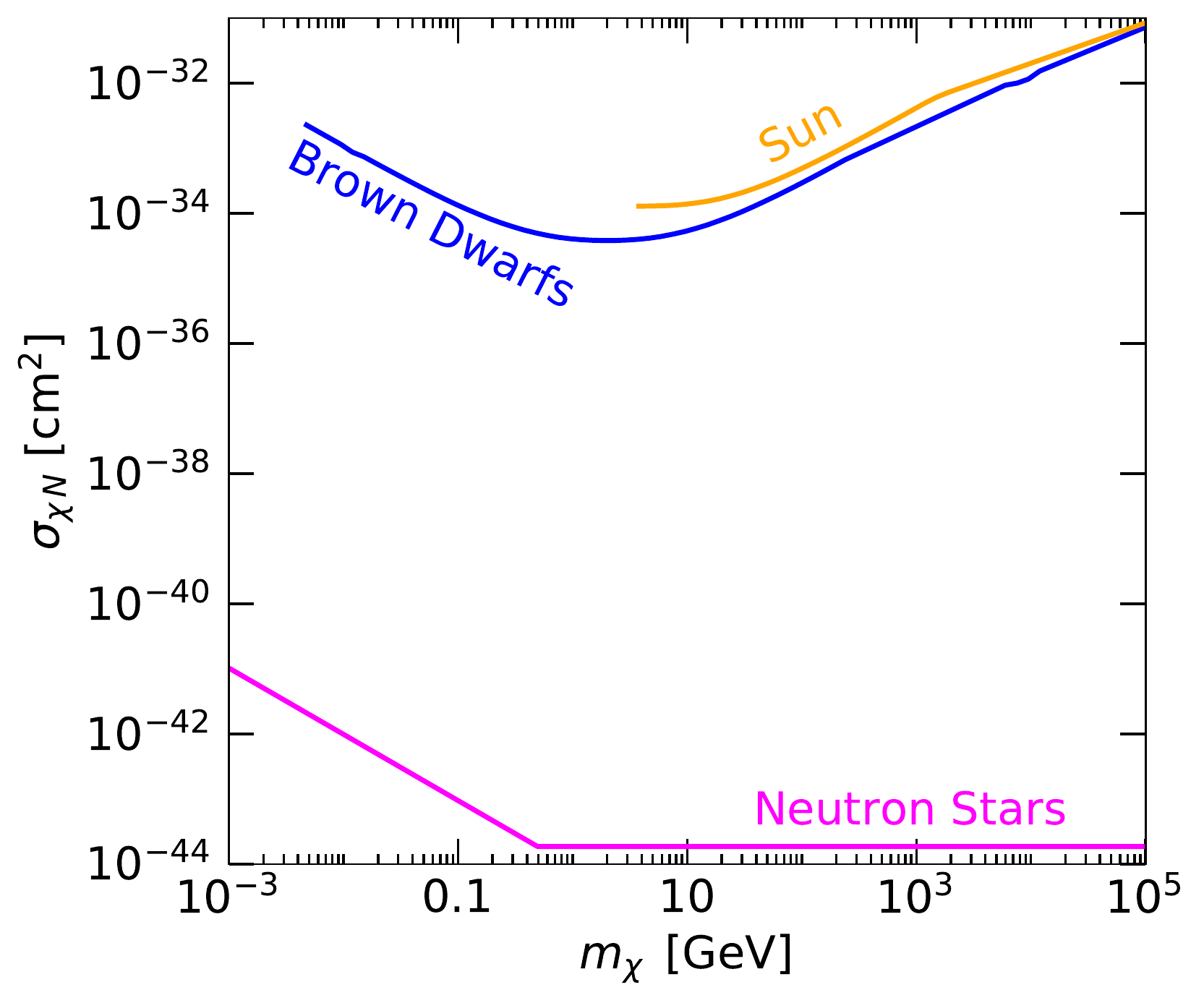}
    \caption{Comparison of the approximate cross sections producing efficient capture (at 99\% of maximum capture rate) at the local position for the Sun, Brown Dwarfs, and Neutron Stars, as a function of DM mass. Cross sections above these values produce comparable DM annihilation rates. We show a benchmark Brown Dwarf with radius that of Jupiter, and mass 0.0378$\,M_{\odot}$. The neutron star benchmark has a radius of 10 km and mass 1.4$\,M_{\odot}$.}
    \label{fig:objects}
\end{figure}

Figure \ref{fig:objects} shows the approximate cross sections at which capture becomes efficient for different celestial bodies at our local position. For definiteness, we plot contours  corresponding to 99\% efficient capture (i.e. $C=0.99 C_{\rm max}$).  We emphasize, however, that significant capture rates can also be achieved for lower cross sections.  For example, scattering cross sections an order of magnitude below this line still yield capture rates of $\sim$50\% the maximum capture rate. For the ``Brown Dwarfs" and ``Sun" sensitivities, we calculate an approximate sensitivity by assuming that Brown Dwarfs and the Sun are composed of 100$\%$ hydrogen. The lower end of each DM mass sensitivity curve is truncated by evaporation of DM out of the systems, which significantly curtails the annihilation signal. For the neutron star rates, we note that nuclear effects are not taken into account for this approximation (see Ref.~\cite{Bell:2020obw} for discussion of how this may weaken rates for DM interaction choices). 

Note that the cross sections in Fig.~\ref{fig:objects} apply to both spin-dependent and spin-independent interactions, as there is no coherent enhancement considered for these objects (although Brown Dwarfs and the Sun contain some helium, this is sub-dominant and has been neglected). In the case that only spin-dependent scattering with neutrons is applicable, the only sensitivity would arise from neutron stars (the other systems predominately contain hydrogen, and therefore only protons), and visa versa for spin-dependent proton scattering only. 

It is important to note that while Fig.~\ref{fig:objects} shows the cross sections corresponding to $99\%$ efficient capture in the given object, the maximum capture rates themselves also differ between these types of objects. That is, larger radii generally lead to more DM passing through the object, so large objects can efficiently capture DM. On the other hand, this will also depend on the DM velocity in the system, which when slow can be advantageous as the effective capture radius grows in size. The translation between maximum capture rate, and the cross section it corresponds to, depends on the density of the object. As NSs are the most dense, they lead to the greatest reach in scattering cross section; weaker interactions are more likely occur in a denser material. This means that while NSs have a superior reach in cross section, they do not necessarily provide the largest capture rate, and therefore do not necessarily provide the largest annihilation rate when scattering/annihilation equilibrium is reached. The relative sizes of capture rates will be compared for the Milky Way environment in Sec.~\ref{sec:mw}.

At masses below a few GeV, the Sun no longer provides any sensitivity to compact-body focused annihilation, due to the efficient evaporation of any captured DM particle. However, NSs and BDs continue to provide significant sensitivities. While the neutron star cross section is smaller, if DM has sufficiently large scattering cross sections, BDs may actually provide the dominant signal, as they have (i) higher number densities in the Galaxy, and (ii) larger DM capture rates due to their larger radii. For lower DM cross sections, NSs potentially provide the only sensitivity. For DM masses above 4 GeV with sufficiently large scattering cross sections, the long-lived particles from solar gamma rays may also be present, depending on the scattering cross section. In that case, very strong limits have already been set~\cite{Leane:2017vag,Albert:2018jwh}. We again stress that the cross sections shown do not correspond to the maximum possible cross-section sensitivity; they instead correspond to scattering cross sections which will approximately provide the largest possible signal for the given object. Therefore, cross sections smaller than those shown can still be probed, given sufficient telescope sensitivity to the smaller signals that would be produced from the lower cross sections.

\subsection{Dark Matter Annihilation in Celestial Bodies}
\label{sec:DM_NS}
 
Once a DM particle becomes trapped in a BD or NS, it has two possible fates. In the case that DM annihilation is forbidden (for example, in the well-studied case of asymmetric DM) the DM density will build-up near the core, potentially leading to eventual black hole formation and collapse. On the other hand, in cases where DM can self-annihilate, there is an interplay between the capture and annihilation in the NS or BD. DM annihilation can deplete the incoming DM density, such that the number of DM particles inside the object $N(t)$ evolves over time, governed by \cite{Kouvaris:2010}
\begin{equation}
\frac{dN(t)}{dt} = C_{\rm tot} - C_A N(t)^2
\label{eq:evolve}
\end{equation}
where $C_{\rm tot}$ is the total capture rate given in Eq.~\ref{eq:multiscatter_total} and $C_A = \langle\sigma_A v\rangle/V$ is the thermally averaged annihilation cross section over the effective volume in which the annihilation takes place. Eq.~\ref{eq:evolve} has the well known solution
\begin{equation}
    N(t) = \sqrt{\frac{C_{\rm tot}}{C_A} }\tanh \frac{t}{t_{\rm eq}},
\label{eq:N evolution}
\end{equation}
where $t_{\rm eq} = 1/\sqrt{C_A C_{\rm tot}}$ is the timescale over which the equilibrium between capture and annihilation of DM within the object is reached. Under the equilibrium condition, the annihilation rate ($\Gamma_{\rm ann}$) is simply:
\begin{equation}
    \Gamma_{\rm ann} = \frac{\Gamma_{\rm cap}}{2} = \frac{C_{\rm tot}}{2},
\label{eq:equlbm}
\end{equation}
where the factor of 2 comes from the fact that in each annihilation event, 2 DM particles are involved. We note from Eqs.~\ref{eq:multi scatter} and \ref{eq:equlbm} that if equilibrium between capture and annihilation is reached, the annihilation rate is proportional to the local DM density i.e. $\Gamma_{\rm ann} \propto n_{\chi}$. The rate will also be proportional to the number density of neutron stars in that region, so the total annihilation rate via BD or NS capture is $\Gamma_{\rm ann} \propto n_{\chi} n_{\rm BD/NS}$. 
\subsection{Dark Matter Annihilation to Long-Lived Mediators}

If DM is captured by, and subsequently annihilates within, celestial objects, several outcomes are possible depending on the annihilation products.  If DM annihilates promptly into SM final states, they will be absorbed in the material of the celestial body in which they were created, heating it \cite{Kouvaris:2007ay,
Bertone:2007ae,
deLavallaz:2010wp,
Kouvaris:2010vv,
Guver:2012ba,
McCullough:2010ai,
Baryakhtar:2017dbj,
Raj:2017wrv,
Bell:2018pkk,
Garani:2018kkd,
Chen:2018ohx,
Dasgupta:2019juq,
Hamaguchi:2019oev,
Camargo:2019wou,
Bell:2019pyc,
Garani:2019fpa,
Acevedo:2019agu,
Joglekar:2019vzy,
Joglekar:2020liw,
Bell:2020jou,
Dasgupta:2020dik,
Garani:2020wge,Leane:2020wob}.  However, models of hidden-sector DM provide another possibility whereby DM annihilates into SM-neutral meta-stable particles.  These ``mediators'' ultimately decay to SM particles, but can be long-lived due to weak coupling and/or approximate symmetries.  In some models, they may also be produced with a substantial Lorentz boost $\eta$. These features allow the mediator to escape the celestial object and then decay in vacuum.  The products of these mediator decays are then observable through searches closely related to the standard indirect detection searches for halo annihilation.

In order to calculate the sensitivities for possible signals, we assume that the mediator $\phi$ has a sufficiently long lifetime $\tau$ or a sufficiently large boost factor $\eta \approx {m_{\chi}}/{m_\phi}$ such that the decay length $L$ exceeds the radius of the object $R$, as
\begin{equation}
L = \eta \beta \tau \simeq \eta c \tau > R.
\label{eq:decay length}
\end{equation}
The differential energy flux (henceforth referred to simply as ``energy flux") at Earth from long-lived particles in celestial bodies is given by~\cite{Leane:2017vag}
\begin{equation}
E^2\frac{d\Phi}{dE} = \frac{\Gamma_{\rm ann}}{4\pi D^2}\times E^2\frac{dN}{dE} \times \mathrm{BR(X\rightarrow SM)} \times P_{\rm surv},
\label{eq:flux}
\end{equation}
where $D$ is the distance to Earth, $\mathrm{BR(X\rightarrow SM)}$ is the branching ratio of the mediator to a given SM final state. The probability of the signal surviving to reach the detector near Earth, $P_{\rm surv}$, provided the decay products escape the object is~\cite{Leane:2017vag}
\begin{equation}
    P_{\rm surv} = e^{-R/\eta c \tau} - e^{-D / \eta c \tau},
\end{equation}
where $R$ is the object's radius. In order to estimate the sensitivities for signals in our analysis, we further assume that the decay of the mediator does not significantly alter the morphology of the annihilation signal (compared to direct annihilation into standard-model particles). This can be accomplished in two ways, either: (1) the mediator decays reasonably close to the source, (2) the mass splitting between the DM particle mass and mediator mass is much larger than the mass splitting between the mediator mass and the mass of the standard model particles it decays into (i.e., it is very boosted). However, as long as the decay impact parameter is short compared to the (Galaxy-scale) distances over which the BD/NS and DM density profiles are varying, this will not significantly impact the results.

We also assume that the mediator escapes the object without attenuation. This assumption is generally reasonable when the mediator particle is long-lived due to its weak coupling with SM particles -- which tend to also suppress scattering cross sections.
For example, assuming that the same coupling $g'$ controls both decays and scatters of the mediator off protons, the expected inverse path-length for decays (in the celestial body's rest frame) scales as $\Gamma_{\rm decay} \sim g'^2 m_\phi/\eta$, where $\eta$ is a boost factor.  Meanwhile, the expected rate for scattering scales as $\Gamma_{\rm scatter} \sim g'^2 \alpha \mu^2/s^2 n \sim g'^2 \alpha / (\eta m_\phi)^2$, where $\alpha$ is an SM coupling constant $\lesssim 1/10$, $n$ the number density of SM matter, and $\mu$ is the DM-proton reduced mass.  Therefore, even within dense compact objects such as NSs with $n\sim (100 \,{\rm MeV})^3$, decays are the dominant means of attenuation so long as 
\begin{equation}
\frac{\Gamma_{\rm scatter}}{\Gamma_{\rm decay}}\simeq \frac{\alpha n}{\eta  m_{\phi}^3 } \simeq \frac{\alpha}{\eta} \left(  \frac{100\, \mathrm{MeV}}{m_{\phi} }\right)^3.
\end{equation}
For mediators heavier than $1$ MeV and/or produced with appreciable boost, this is typically $<1$.  Attenuation by scattering is even less relevant in BDs, due to their much lower densities. 


\subsection{Dark Matter Annihilation in the Halo}
\label{sec:DM_halo}
Particle DM models relevant for BD/NS-focused annihilation can also, in general, produce the more standard signal of DM particles annihilating in the halo which hosts the BD/NS population.  To facilitate future comparisons of the two signals, we briefly review the standard halo annihilation rate and highlight important contrasts with the  BD/NS-focused annihilation rate.  

In particular, the standard halo annhilation rate scales quadratically with DM density ($\propto n_{\chi}^2$), while BD/NS annihilation rate scales linearly with $n_{\chi}$ as seen in Eqs.~\ref{eq:tot_capture} and \ref{eq:equlbm}. Therefore the expected signals from BD/NS focused annihilation will be different from the standard halo annihilation signals. The annihilation rate in the halo scales as
\begin{equation}
    \Gamma_{\rm halo} \propto \frac{\langle\sigma_A v\rangle\, n_{\chi}^2}{2},
\end{equation}
which highlights the characteristic scaling that is proportional to the thermally-averaged annihilation cross section and the square of the number density of DM particles. The quadratic dependence of the halo annihilation rate on the DM density implies that the brightest annihilation targets typically correlate with peaks in the DM density, such as the Milky Way Galactic center, the centers of dwarf galaxies, and of distant galaxy clusters. 

In general, the annihilation cross section can be expanded in velocity ($v$) as 
\begin{equation}
    \langle\sigma_A v\rangle \propto v^\ell,
\label{eq:v-expansion}
\end{equation}
where the leading rate is found when $\ell = 0$, i.e. an $s-$wave contribution is present. The next leading term in velocity is the $p-$wave contribution (with $\ell = 2$). 

From the Boltzmann velocity distribution, $\langle v\rangle \sim \sqrt{T}$ so that 
\begin{equation}
    \langle\sigma_A v\rangle \propto x^{-n}
\label{eq:v-expansion-x}
\end{equation}
where $x = m_{\chi}/T$ and $n = p/2$. Using this expansion, the WIMP relic density can be estimated as \cite{Kolb:1990vq}
\begin{equation}
    \Omega h^2 = 0.0845 \, (n + 1) x_f^{n+1} \sqrt{\frac{100}{g_{\star}}} \left( \frac{10^{-10} \mathrm{GeV^{-2}} }{\langle\sigma v\rangle_0} \right),
\end{equation}
where $x_f$ is the freeze-out time and $g_{\star}$ is the number of degrees of freedom at freeze-out. Given the present day DM density, and assuming an $s$-wave dominant annihilation rate, we obtain a thermal annihilation cross section of \mbox{$\langle\sigma_{\rm ann} v\rangle_{\rm s-wave} \sim 2.2 \times 10^{-26}$ $\mathrm{cm}^3 \mathrm{s}^{-1}$~\cite{Steigman:2012nb}.} 

Importantly, we note that for a $p$-wave dominated annihilation rate, the annihilation cross section today will be velocity suppressed $\langle\sigma v\rangle_{\rm p-wave} \propto v^2$. This means that the expected cross section for typical DM velocities of $\sim$200 km/s today will be about 10$^{-5}$ times smaller than expected for s-wave annihilation. Noting that cutting-edge experiments are only beginning to probe the fluxes expected from s-wave annihilation processes, we stress that $p$-wave rates are typically unobservable in the halo. 

By contrast, for celestial body focused annihilation, DM annihilation typically occurs deep within the focusing object, when myriad captures have produced a sharply peaked DM density. In this case, DM can annihilate efficiently even when the annihilation cross section is extremely low. Smaller DM cross sections simply correspond to a longer equilibration timescale, rather than a smaller DM signal. 

To summarize, halo-based annihilation is quadratically dependent on the DM density, and linearly dependent on the DM annihilation rate. Celestial-body focused annihilation that has reached equilibrium is linearly dependent on the DM density has a flux that depends on the scattering rate rather than the DM annihilation rate. These differences provide two stark observable signatures that can differentiate halo and celestial-body focused annihilation.

\section{Milky Way Galactic Center Signal}
\label{sec:mw}

We first investigate the detectability of our BD/NS-focused annihilation signal in the Milky Way's Galactic center, where the luminosity of the signal is expected to be high due to the large population of NSs and BDs in the inner parsecs of the galaxy. In this section we introduce specific models for the (i) DM velocity distribution, (ii) NS number density, and (iii) BD number density in the inner galaxy.  These, together with the DM density (modeled as a generalized NFW profile as in Eq.~\ref{eq:DM_density}, determine the GC BD/NS-focused annihilation fluxes for a given capture rate. We will then compare the resulting BD/NS-focused annihilation fluxes to both halo annihilation fluxes and telescope sensitivities in this complex region.

\subsection{Modeling Milky Way Velocity Components}
In addition to the DM density, the DM velocity dispersion strongly affects the rate at which DM particles in the vicinity of a NS fall into its potential well and intersect the NS surface. We calculate the DM velocity dispersion using models for the mass distribution and velocity profile of the Milky Way following Ref.~\cite{Sofue:2013kja}. This model assumes five components for the total mass $M(r)$: the central Black Hole (BH) with mass $M_{\rm BH}=4 \times10^6$ M$_{\odot}$, an exponential disk ($\rho_{disk}$), an inner and outer spheroidal bulge ($\rho_{inner}$ and $\rho_{outer}$) and a DM generalized NFW halo as per Eq.~\ref{eq:DM_density} ($\rho_{DM}$). Our DM density profile is normalized to the local DM density of $0.42$ GeV/cm$^3$, and the inner slope is taken to be either
$\gamma = 1.0$ or $\gamma =1.5$, and the scale radius is chosen as $r_s =$ 12 kpc (these values are our DM density profile choices, not adapted from Ref.~\cite{Sofue:2013kja}). The steeper choice for the inner profile slope can be motivated by expectations from adiabatic contraction in the inner Galaxy~\cite{gnedin2011halo,DiCintio:2014xia}.

These components are combined to provide a model for the total mass,
\begin{align}
    M(r) &= M_{\rm BH} \nonumber\\
    &+4\pi \int_0^r (\rho_{\rm outer} + \rho_{\rm inner} + \rho_{\rm disk} + \rho_{\rm DM}) dr \, .
\label{eq:Milky_Way_Mass}
\end{align}
 From this mass distribution, it is straightforward to calculate the galactic rotation velocity, as per Ref.~\cite{Sofue:2013kja}. However, it is important to note that the models for velocities towards the inner Galaxy are not robust. Indeed, recent work finds significantly lower Galactic velocities in the inner $\sim3~$kpc~\cite{Chemin_2015}. Such lower velocities would substantially boost our expected DM capture rates, as the lower velocities allow the DM to be more easily captured via gravitational focusing. However, to be conservative, we do not consider these lower velocities. We note that the circular orbital velocities $v_c$ of Ref.~\cite{Sofue:2013kja}, are related to the velocity dispersion $v_d$ by $v_d=\sqrt{3/2} v_c$.

\subsection{Neutron Star Population in the Galactic Center}

We now investigate the properties of the NSs that are relevant for our GC signal. There is strong evidence for a population of NSs near the Milky Way GC. In particular, observations of hundreds of O/B stars currently located in the central parsec indicate a high rate of \textit{in situ} NS/BH formation in this region~\cite{2003:Genzel,2003:Levin}. Indeed, it is expected that a dense system of compact objects reside in the GC region, and the expected population has been estimated in the literature~\cite{Freitag:2006qf,Hopman:2006xn,Merritt:2009mr,Generozov:2018niv}\footnote{For arguments to the contrary, we note that the observation of radio pulsars near the galactic center has proven unexpectedly difficult, leading some to conclude that there is a ``missing pulsar problem" which may indicate an unexpected absence of pulsars near the Galactic center~\cite{Dexter:2013xga}. However, Ref.~\cite{Chennamangalam:2013zja} argues that this is merely an observational effect, and the pulsar density near the Galactic center is still likely to be large.}. The number densities of black holes and neutron stars in the Galactic Center region have been previously obtained with numerical simulations of nuclear star cluster dynamics~\cite{Generozov:2018niv}.  These studies utilize the Fokker-Planck equation to numerically evolve the radial distribution of stars and compact objects over time, taking into account two-body relaxation. 

In Ref.~\cite{Generozov:2018niv}, two types of nuclear cluster models were described. One is the `Fiducial $\times$ 10' model where it is assumed that compact objects which are injected near the present disk of massive stars at $\sim$ 0.3 pc will diffuse outwards via two-body scattering. The formation rates of NS and BH in this model are respectively taken to be $\dot{N}_{\mathrm{NS}} = 4 \times 10^{-5}$ yr$^{-1}$ and $\dot{N}_{\mathrm{BH}} = 2 \times 10^{-5}$ yr$^{-1}$ corresponding to the present day formation rates of massive stars. This model also takes into account `Primordial' NS's of masses 1.5 $M_{\odot}$ which are deposited impulsively at $t = 0$. The other model is labeled the `Fiducial' model, and utilizes very conservative star formation rates (SFR) of $\dot{N}_{\mathrm{NS}} = 4 \times 10^{-6}$ yr$^{-1}$ and $\dot{N}_{\mathrm{BH}} = 2 \times 10^{-6}$ yr$^{-1}$, which are approximately an order of magnitude lower than the present day star formation rate. The order of magnitude smaller formation rates for the `Fiducial' model in Ref.~\cite{Generozov:2018niv} were motivated by the results of Ref.~\cite{2011:Pfuhl} where it was found that that the SFR $1-5$ Gyr ago was $1-2$ orders of magnitude smaller than the present day rate. However, Ref.~\cite{2011:Pfuhl} only took into account low-mass stars ($\lesssim$ 2 M$_{\odot}$), and their results do not directly constrain the rate of NS/BH formation within the star-forming discs if the top-heavy disc IMF is truncated below a few solar masses (for a more detailed discussion see Ref.~\cite{Generozov:2018niv}). Therefore, given the observational uncertainties, both the `Fiducial' and `Fiducial $\times$ 10' models outlined in Ref.~\cite{Generozov:2018niv} are potentially equally good candidates for representing a generic NS distribution in the nuclear star cluster. 

The NS number density for the `Fiducial $\times$ 10' model at 10 Gyr is roughly a factor of $3-4$ times higher than that of the `Fiducial' model (see Fig. 2 in Ref.~\cite{Generozov:2018niv}). In this work, we use the `Fiducial $\times$ 10' model to demonstrate our idea, while noting that the signals with the `Fiducial' model will be roughly a factor of $3-4$ smaller. For simplification, we assume all NSs have a mass of 1.5 M$_{\odot}$. We also note that other studies focused on modeling the compact object distribution in the nuclear star clusters \cite{Freitag:2006qf,2016:Aharon} are in rough agreement with the NS number density estimates of Ref.~\cite{Generozov:2018niv}.

For a radial distribution for NSs, we extract the NS radial number density distribution of the `Fiducial $\times$ 10' model shown in Fig. 2 of ~\cite{Generozov:2018niv}. This provides a NS number density,
\begin{eqnarray}
\begin{aligned}
n_{\rm NS} &= 5.98 \times 10^3 \left(\frac{r}{1 \mathrm{pc}} \right)^{-1.7} \mathrm{pc^{-3}}; \ \ \ 0.1 \,\mathrm{pc} < r < 2 \,\mathrm{pc} , \nonumber \\
&= 2.08 \times 10^4 \left(\frac{r}{1 \mathrm{pc}} \right)^{-3.5} \mathrm{pc^{-3}}; \ \ \ \ \ \ \ \ r > 2 \,\mathrm{pc}.
\label{eq:ns_density}
\end{aligned}
\end{eqnarray}

\subsection{Brown Dwarf Population in the Galactic Center}

A huge number of BDs are expected to be present in the Milky Way. In Ref.~\cite{2017:Muzic}, it was estimated that the Milky Way may contain as many as $25-100$ billion BDs. To obtain the radial distribution of BDs, we use the BD distribution function outlined in Refs.~\cite{2013:Kroupa,Amaro-Seoane:2019umn}. In this treatment, the Kroupa Initial Mass function (IMF) \cite{2013:Kroupa} is extended to include sub-stellar BD masses. The BD IMF is described by a broken power law of the form $\frac{dN_{\mathrm{BD}}}{dm} \propto m^{-\alpha}$, where $N_{\mathrm{BD}}$ and $m$ are the BD number and mass respectively, and $\alpha = 0.3$ \cite{2013:Kroupa}. The BD number density in the mass range $0.01-0.07$ M$_{\odot}$ is given by~\cite{2013:Kroupa,Amaro-Seoane:2019umn}
\begin{equation}
     n_{\rm BD} = 7.5 \times 10^{4} r_{\mathrm{pc}}^{-1.5} \mathrm{pc^{-3}},
 \end{equation}
where $r_{\mathrm{pc}}$ is the radius of the containment volume in parsecs. Unlike many NSs and BHs, BDs are not born with natal kicks. 3-Body interactions in the dense Galactic center might eject some BDs from the center, but that number is expected to be small. 

Note that for our BD calculations, we take the average mass M$_{\mathrm{BD}}$ = 0.0378 M$_{\odot}$ to be representative of the population mass between $0.01-0.07$ M$_{\odot}$, with the mass distribution given by the Kroupa IMF discussed above. We have checked that the error introduced in the total capture rate by using the average mass compared to using the full mass distribution ($\propto m^{-0.3}$) is less than 10\%.

\subsection{Celestial-Body Focused vs. Standard Halo Annihilation}

\begin{figure} [t!]
    \centering
    \includegraphics[width=\columnwidth]{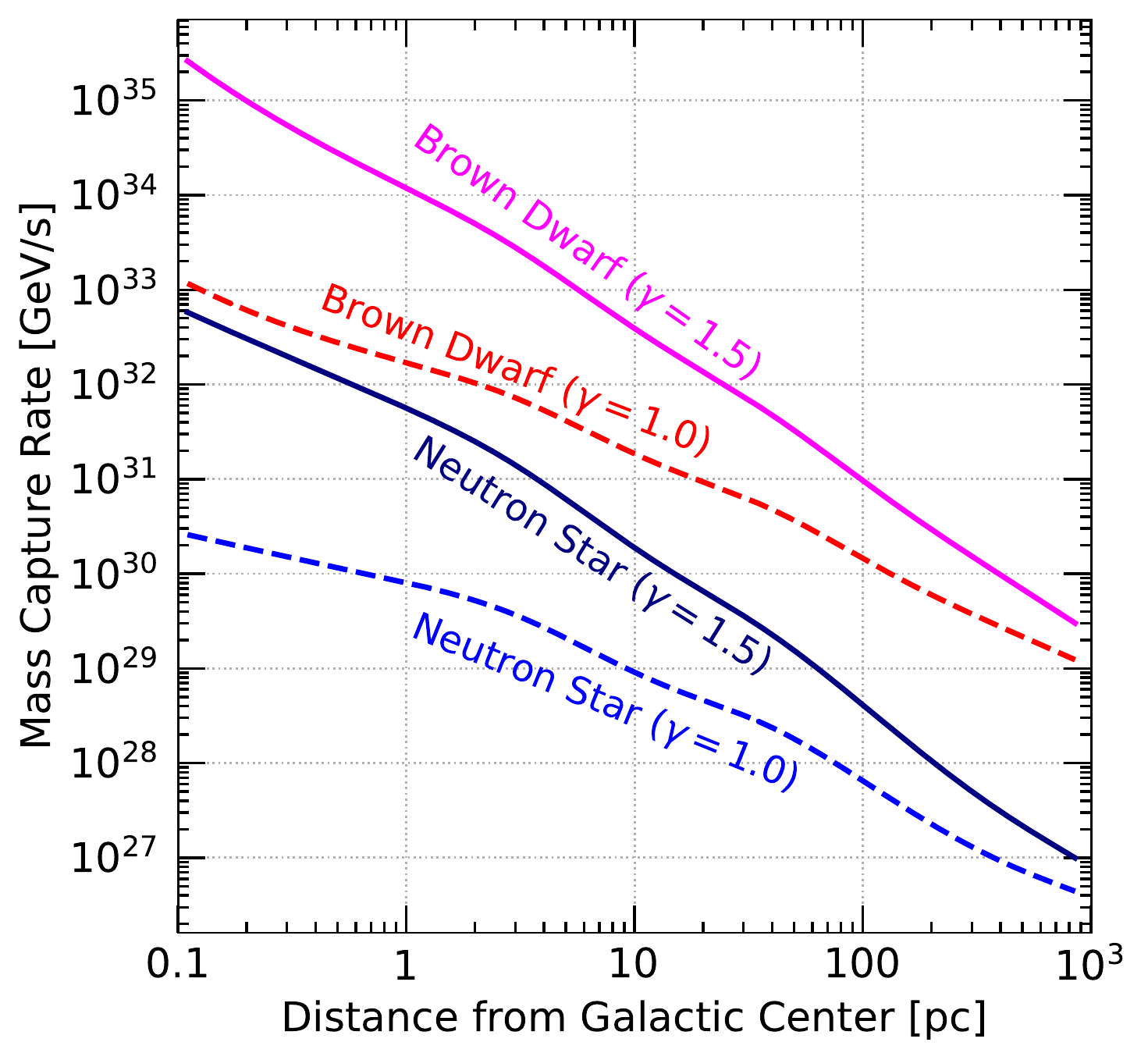}
    \caption{Maximum DM mass capture rates for a single neutron star or brown dwarf as a function of radius ($r$) from the Galactic center. Results are shown for NSs with $R_{\rm NS} = $ 10 km and $M_{\rm NS}$ = 1.4 $M_{\odot}$, and BDs with $M_{BD}$ = 0.0378 $M_{\odot}$ and $R_{BD}$ = $R_J$ (where $R_J$ is the radius of Jupiter). We show varied results for generalized NFW DM profiles, with $\gamma = 1.0,\,1.5$.}
    \label{fig:maxcap}
\end{figure}

We first calculate the DM capture rate from a \textit{single} NS or BD located at a distance $r$ from the Galactic Center. To do this, we use the multi-scatter formalism outlined in Sec. \ref{sec:capture}, taking the DM density and Galactic velocity dispersion as defined in the previous subsection.

Figure~\ref{fig:maxcap} shows the NS and BD capture rates as a function of radius $r$, assuming a maximum capture rate for DM particles. These capture rates correspond to a \textit{single} BD/NS that accumulates DM particles with any scattering cross section that is larger than the cross sections shown in Fig.~\ref{fig:objects}. We see that BDs have a much larger maximum capture rate than NSs. This is because the maximum capture rate is determined by the total DM flux that passes through the object; the effective capture radius is larger for BDs, because their radius that is about 1,000 times larger than NSs. The wiggles in the plot are due to the interplay between the DM density and halo velocity. For demonstration, we show two cases of the NFW slope, $\gamma = $ 1.0 (standard NFW) and $\gamma = $ 1.5 (generalized NFW, with a steep slope).  The smaller NFW slope decreases the DM density in the GC region (where the BDs/NSs are present in largest numbers), which leads to a lower total capture rate. 

To calculate the total capture rate from the GC population of NSs, $C_{\rm tot, NS}$,  we use Eq.~\ref{eq:tot_capture}, and the number density of neutron stars $n_{\rm NS}$ from Eq.~\ref{eq:ns_density}. We integrate over a volume between $r =$ 0.1 pc to $r =$ 100 pc. The cutoff radius of $r = 0.1$ pc is chosen because the DM cusp-like profile might break down at lower radii. The outer radii of 100 pc is chosen because outside of this region, NS/BD number density too low to substantially change our results. Because NSs can receive substantial natal kicks due to asymmetries in the supernova explosion mechanism, only about 60\% of the NS that are born near the Galactic center are expected to be retained within our volume~\cite{Generozov:2018niv}. Note that if the NSs in the nuclear clusters receive even stronger natal kicks such that $\sim$ 90 \% of all NSs are ejected out \cite{Panamarev:2018bwq}, the signals will correspondingly decrease. For the NFW density slope $\gamma = 1.5$, we obtain a total DM capture rate by all the NSs in the GC region of $C_{\rm NS, GC}$ = $\Gamma_{\mathrm{cap,NS}}$ $\sim$ 6 $\times$ 10$^{36}$ GeV/s. For $\gamma = 1.0$, we obtain a total DM capture rate onto NSs of $\Gamma_{\mathrm{cap,NS}}$ $\sim$ 10$^{35}$ GeV/s.

\begin{figure*}[t!]
    \centering
    \includegraphics[width=\columnwidth]{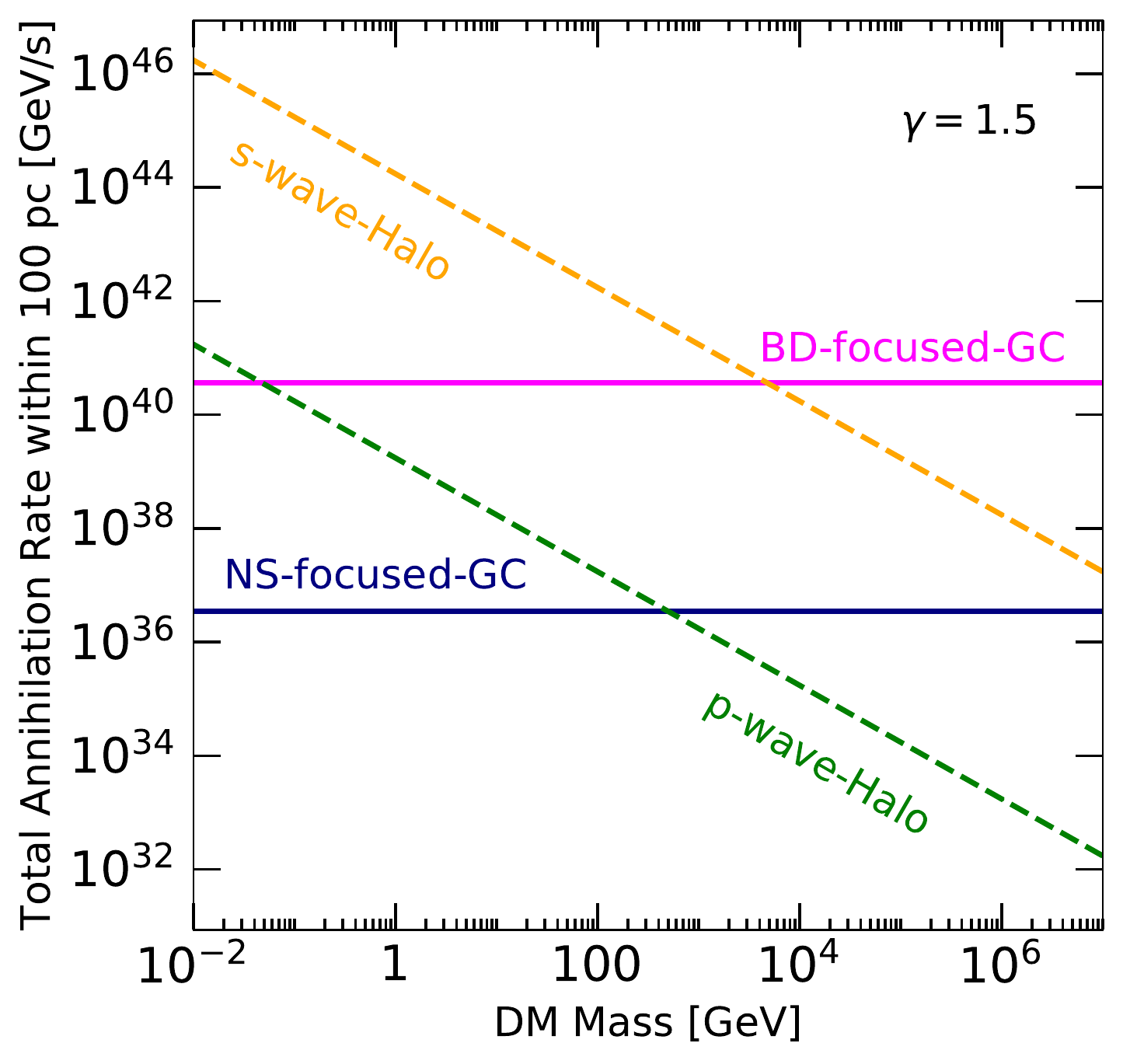}
    \includegraphics[width=\columnwidth]{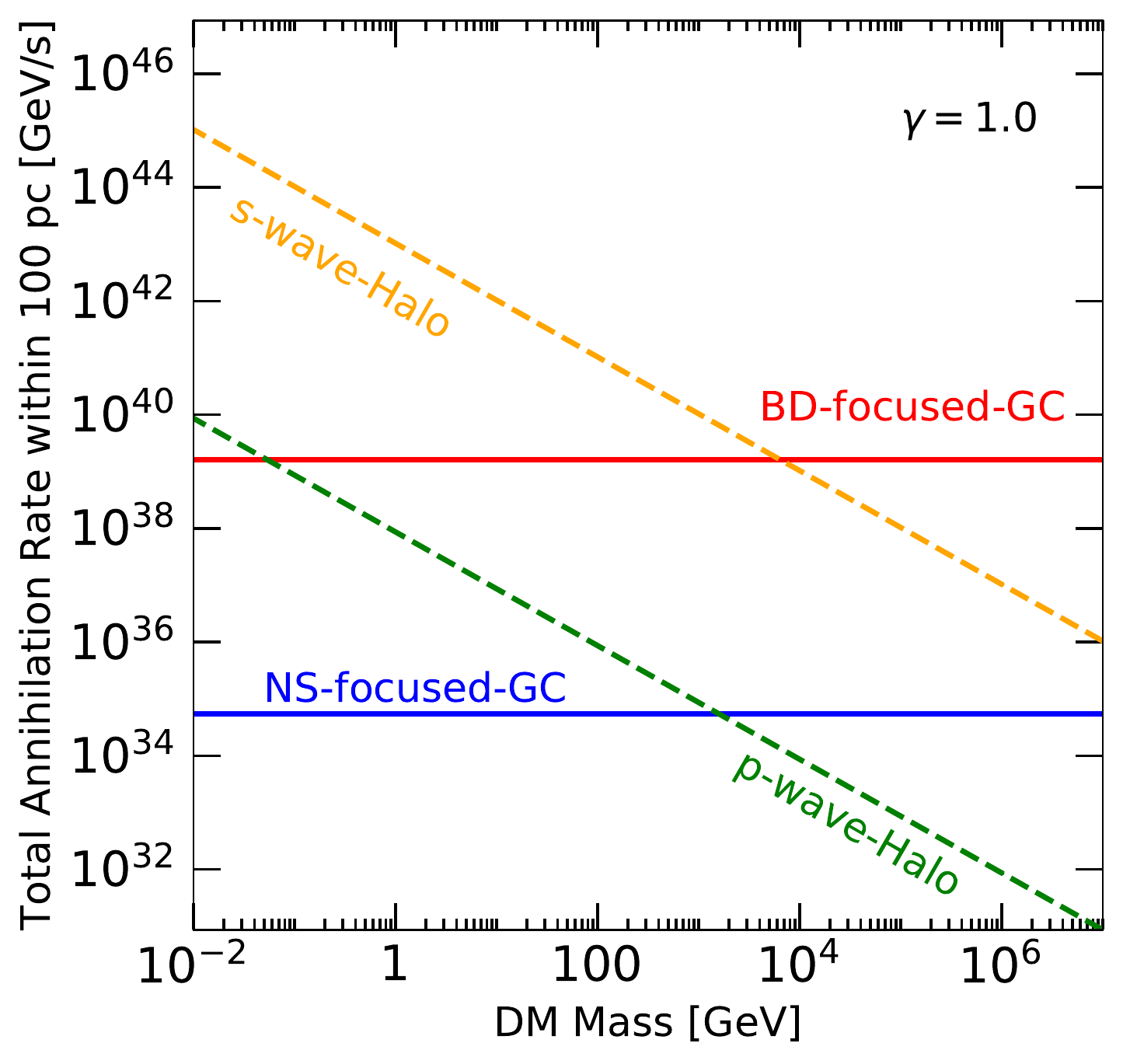}
    \caption{NS-focused annihilation and BD-focused annihilation (solid) vs. halo annihilation in the Milky Way Galactic center, for $s-$wave and $p-$wave DM rates (dashed) for varying DM mass. This plot assumes the maximum capture rates for BD/NS. Two panels for different NFW slope $\gamma$ = 1.0 and 1.5 are shown.}
    \label{fig:rates}
\end{figure*}

To calculate the total capture rate from the GC population of BDs,
we follow the same integration procedure as NSs (but note that BDs do not receive natal kicks capable of ejecting them from the GC). We find that the total capture rate from the GC population of BDs is $\Gamma_{\mathrm{cap,BD}}$ $\sim$ 7 $\times$ 10$^{40}$ GeV/s for $\gamma =$ 1.5. For $\gamma$ = 1.0, the BD capture rate is $\Gamma_{\mathrm{cap,BD}}$ $\sim$ 3 $\times$ 10$^{39}$ GeV/s.  

For a celestial object in equilibrium, the total annihilation rate from all NSs or BDs corresponds to half the total capture rates (because self-annihilation removes two DM particles), as shown in Eq.~\ref{eq:equlbm}. If this entire flux escapes the celestial body through annihilation into the long-lived mediator and then decays, the total annihilation rate within 100 pc from BDs will be $\Gamma_{\mathrm{ann},\mathrm{BD}}$ = $3.5 \times 10^{40}$ GeV/s for $\gamma$ = 1.5 and $\sim 1.6 \times 10^{39}$ GeV/s for $\gamma = 1.0$. For NSs, the total annihilation rate will be $\Gamma_{\mathrm{ann,NS}}$ = 3 $\times$ 10$^{36}$ GeV/s for $\gamma = 1.5$ and $\sim$ 5 $\times$ 10$^{34}$ GeV/s for $\gamma$ = 1.0. 

Figure~\ref{fig:rates} demonstrates the relative strength of the NS-focused and BD-focused annihilation, compared to halo annihilation, as a function of DM mass for both $p$- and $s$-wave DM for $\gamma$ = 1.0 and 1.5. The total halo annihilation rate is calculated by integrating the annihilation rate along the line of sight over the whole angular range of the sky. For NS-focused annihilation, when $m_{\chi} \lesssim 10^3 $  GeV, halo annihilation dominates for both $s-$ and $p-$ wave DM. When $m_{\chi} \gtrsim 10^3$ GeV, NS-focused annihilation becomes dominant over the $p-$wave halo annihilation rate. For BD-focused annihilation, when $m_{\chi}$ $>$ 0.1 GeV the signal is larger than $p-$wave annihilation. For $m_{\chi} \lesssim 3000 $  GeV, halo annihilation dominates for $s-$wave halo signal for BDs. For $m_{\chi} \gtrsim 3000 $ GeV, BD-focused annihilation is dominant to both $s-$ and $p-$ wave signals. This result holds for both $\gamma = $ 1.0 and 1.5.

In the heavy DM case, celestial- body-focused annihilation enhancement is particularly pronounced for both NS and BDs, as their linear dependence on the DM number density means that their flux is constant with DM mass, while the halo rate becomes suppressed.

\section{Dark Matter Parameter Space}
\label{sec:results}

To translate these observations into constraints on the DM parameter space, we now calculate the capture rates and corresponding cross section limits that can be constrained via current $\gamma$-ray observations.

\subsection{Gamma-Ray Telescope Sensitivity}

\begin{figure} [t!]
    \centering
    \includegraphics[width=\columnwidth]{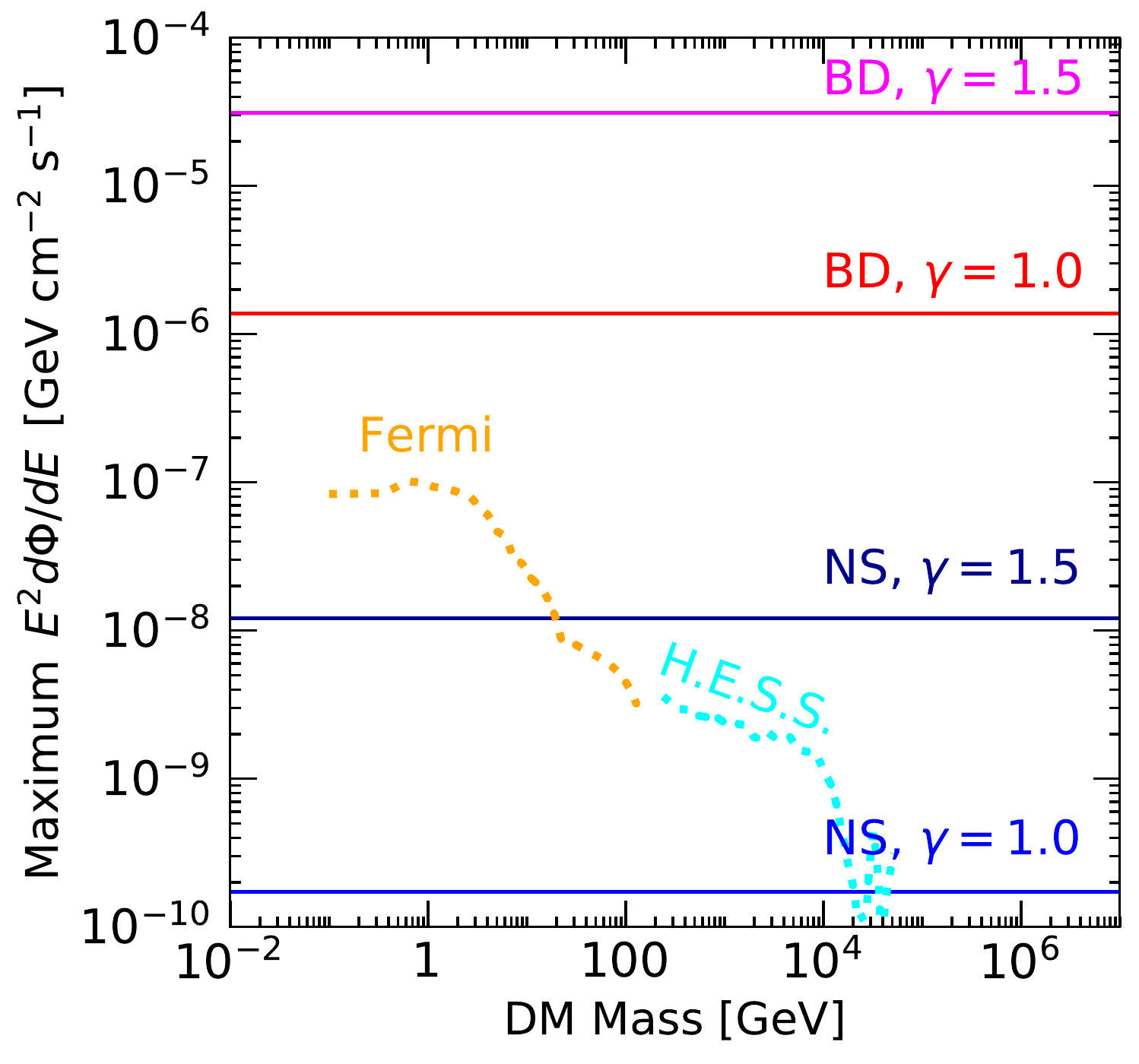}
    \caption{Maximum $E^2\, d\Phi/dE$ values (at $E_\gamma\approx m_\chi$) for the Galactic center population of Brown Dwarfs or Neutron Stars, for DM densities described by NFW profiles with indices of 1.0 or 1.5, as labelled. The Galactic center gamma-ray fluxes measured by \textit{Fermi} and H.E.S.S. are shown for comparison, where $E_\gamma\approx m_\chi$ is assumed (see text for details).}
    \label{fig:gammaflux}
\end{figure}

To set scattering cross section limits, we use the fluxes already measured by \textit{\textit{Fermi}} and H.E.S.S. at the Galactic center~\cite{Malyshev:2015hqa}. We use \textit{Fermi} data for all observations corresponding to DM masses less than $\mathcal{O}(100)~$GeV. This is appropriate because the \textit{Fermi}-LAT has produced sensitive measurements across the entire sky. At higher masses, however, our results require separate instrumentation. For the Sun, we utilize the solar limits derived in Ref.~\cite{Albert:2018jwh} using the HAWC telescope, because atmospheric Cherenkov telescopes (like H.E.S.S. and VERITAS) are not designed to work when pointed at the Sun. For GC limits at TeV energies we utilize H.E.S.S. data, because HAWC and VERITAS lie in the northern hemisphere, and have poor exposures of the Galactic center region. 

We set limits by simply requiring that the normalization of the DM flux does not exceed $100\%$ of the measured gamma-ray flux. This is done by determining the smallest scattering cross section for which the energy flux found in Eq.~\ref{eq:flux} (and using Eqs.~\ref{eq:tot_capture} and \ref{eq:equlbm}) exceeds the measured flux in any energy bin. This is a very conservative approach. 

The photon energy spectrum from DM annihilation, and hence the observational constraints we consider, depend on the final states into which DM annihilates. We focus below on modes $\chi\chi\rightarrow\phi\phi$, $\phi\rightarrow2\,\gamma$, where the mediator $\phi$ escapes the system of interest, as per Eq.~\ref{eq:decay length}, before it decays.  This process is expected to dominate over production of only one mediator, due to phase space suppression. The mediator mass and its precise lifetime have little effect on the signal, so long as the mediator lifetime is $\lesssim$ parsec-scale.  We will later comment briefly on other decay modes.

To generate our gamma-ray energy spectra, we use \texttt{Pythia}~\cite{Sjostrand:2007gs}. We create an effective resonance with energy $2\,m_\chi$, by colliding two back-to-back neutral beams. This resonance is then decayed into two mediators, which decay to SM particles. These SM particles can radiate further particles, decay themselves, or shower. All such possibilities are taken into account, and we use the fully decayed spectra in vacuum. 

 \begin{figure*} [t!]
    \centering
    \includegraphics[width=2\columnwidth]{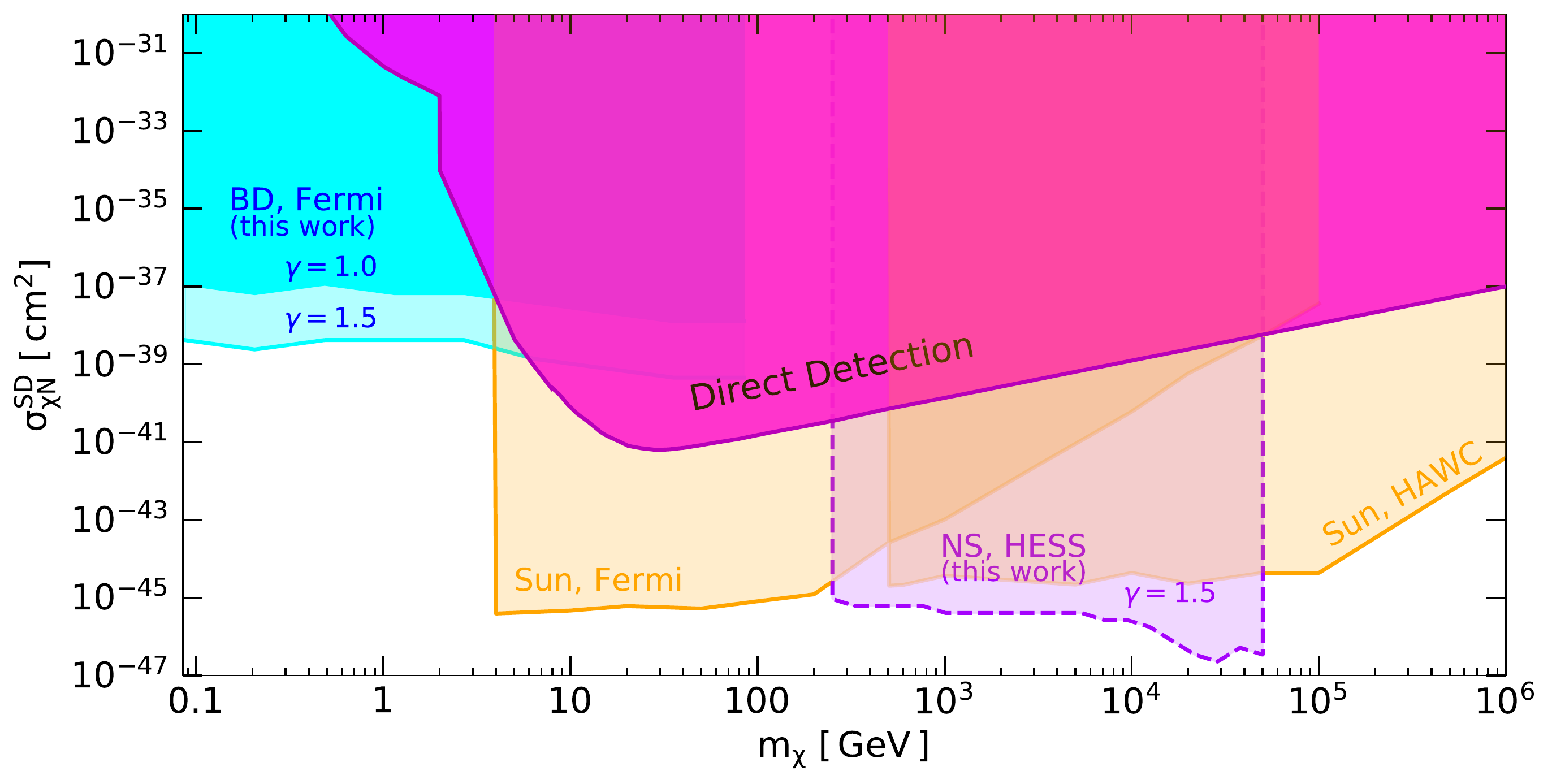}
    \caption{Scattering cross section limits for DM annihilation to long-lived mediators decaying to $\gamma\gamma$ in Brown Dwarfs (using \textit{Fermi}), the Sun (using \textit{Fermi} and HAWC~\cite{Leane:2017vag,Albert:2018jwh}), and Neutron Stars (using H.E.S.S.). The BD and NS limits are new calculations in this work, calculated using the full Galactic Center population of BDs or NSs. The $\gamma=1.0,\,1.5$ values correspond to the inner slope of the DM density profile. Our limits have some assumptions; see text for details.}
    \label{fig:limits}
\end{figure*}

Figure~\ref{fig:gammaflux} shows the relative sizes of the $E^2\, d\Phi/dE$ flux (as defined in Eq.~\ref{eq:flux}) values for direct decay to gamma rays, for BDs and NSs in DM NFW profiles with indices of 1.0 and 1.5. We see that compared to Fig.~\ref{fig:rates} (which only considers the total integrated flux), the maximum values of the fluxes are higher. This is because the energy spectrum for direct decay into gamma rays, $E^2 dN/dE$, is peaked near the DM mass, putting a large amount of gamma rays into a particular energy bin. We have shown the maximum flux value of the whole spectrum; as this is what will generally set the limit relative to the measured telescope flux. For comparison, we show the fluxes measured in the Galactic center by \textit{Fermi} and H.E.S.S., where we have taken for demonstration purposes $E_\gamma\approx m_\chi$ just for this plot, which is a valid approximation given the sharp box spectrum expected for direct decay to gamma rays. We see that BD fluxes are higher than that measured by \textit{Fermi} and H.E.S.S., which leads to strong constraints. On the other hand, for NSs, only the $\gamma=1.5$ NS flux clearly exceeds H.E.S.S. data; as such we will only show how this translates into cross-section sensitivity for NSs for this index choice.

\subsection{Cross Section Limits}

Figure~\ref{fig:limits} shows our cross section constraints for mediator decay to gamma rays, via $\chi\chi\rightarrow\phi\phi$, $\phi\rightarrow2\,\gamma$. We show for comparison, limits obtained for long-lived mediators in the Sun from Refs.~\cite{Leane:2017vag,Albert:2018jwh}, as well as direct detection limits~\cite{Gangi:2019zib, Aprile:2019dbj}. Our analysis, based on existing telescope data, can outperform both solar and direct detection limits. The BD limits can outperform existing limits in the same sub-GeV mass range by more than \textit{nine orders of magnitude}. The reasons for such new powerful bounds are (i) direct detection sensitivities are greatly weakened, because lighter recoils fall below the detection threshold, and (ii) BDs have cooler cores and do not evaporate DM with masses above a few MeV, providing new sensitivity to light DM (the Sun on the other hand, has truncated limits around 4 GeV).  BDs also can outperform spin-dependent indirect detection, although in this region our bounds overlap with solar constraints. Neutron stars outperform SD direct detection by $\sim 4-8$ orders of magnitude, in the $\sim200-10^5$ DM mass range. In this mass range, NSs even can potentially outperform the Sun, by $\sim1-2$ orders of magnitude.

We emphasize that these bounds can be weakened in some scenarios. For example, if the DM density profile near the galactic center is shallower, for example in the case of a cored profile, the bounds would substantially weaken. To show a range of cuspy profiles, we show results for BDs with both $\gamma=1.0$ and $\gamma=1.5$. For NSs, we only show $\gamma=1.5$. This is because the maximum gamma-ray flux produced by NSs is only just detectable over the H.E.S.S. background flux. We emphasize again, however, that we have taken a very conservative approach in setting our limits. As such, it is possible that in a less conservative analysis, the NS signal could potentially be probed across a range of DM masses even with a profile index of $\gamma=1.0$ (note that comparing with Fig.~\ref{fig:gammaflux}, $\gamma=1.0$ can just be detected, but only for a very narrow DM mass range). On the other hand, we also note that recent work has argued that the maximum capture rates for NSs can be smaller due to nuclear effects~\cite{Bell:2020obw}, for varying Lorentz interactions. As such, we only show the NS limit range with a dashed line -- in full model-dependent contexts, the limits will likely be contained somewhere within this range. 

We note several ways to construct models that can change the relative strength of these limits. First, we note that the solar and BD limits require proton scattering, while the NS limits require neutron scattering, so in the case that only one coupling is present, the other limits will disappear. We also note that we have assumed that the mediator has a sufficiently long lifetime and/or boost that it escapes the celestial body in question. However, each of these systems shown have differing radii, and as such, if the mediator lifetime or boost were shorter, the Sun, BD or NS limits may disappear, in that order. Furthermore, much \textit{longer} lifetimes may be probed by the GC populations of BDs/NSs compared to the Sun -- a decay length that is much longer than an A.U. suppresses the flux from the Sun, and depending on mediator boost, can also enlarge the angular region that the signal appears to emanate from. In this sense, the BD/NS limits are the most general; they apply to a wider range of decay lengths.

While we have only shown mediator decay to gamma rays $\chi\chi\rightarrow\phi\phi\rightarrow4\gamma$ in Fig.~\ref{fig:limits}, other final states can also be probed. For electron final states, there is some additional sensitivity at lower DM masses with BDs than can be probed with the Sun, however this is only a few GeV improvement, as the electron gamma-ray spectrum is very soft, it peaks outside \textit{Fermi}'s energy range for any lower DM masses. For $b$-quarks or $\tau$ leptons, there is no additional sensitivity with BDs using \textit{Fermi} compared to current constraints from the Sun. The main reason why $b$-quarks or $\tau$ spectral types do not gain new sub-GeV sensitivity is that their softer spectral shapes peak outside \textit{Fermi}'s sensitivity. As such, upcoming MeV telescopes such as AMEGO and e-ASTROGAM could provide strong limits for these additional final states. Note, however, that generically, the direct decay to photons will provide the strongest constraints.

It also is possible to probe final states other than $\phi\rightarrow 2\,{\rm SM}$ with BDs/NSs. For example, constraints could also be set on $\phi\rightarrow 3\gamma$ processes (motivated by light vectors) and or $\phi \rightarrow \phi' +\gamma$ (e.g. a long-lived dipole-type transition between two massive dark sector states). However due to their spectral shape, we expect these will likely produce weaker constraints.

Lastly, we comment on our expectation that the cross sections shown in Fig.~\ref{fig:limits} will lead to equilibrium being reached. Most stars in the Galactic center nuclear star cluster are expected to be very old, potentially older than $\sim$ 5 Gyr \cite{Generozov:2018niv}. Therefore, to check the equilibration timescale, we calculate the minimum scattering cross section for all BD/NS within 100 pc that allows $t_{\rm eq}$ to be less than $\mathcal{O}$(1 Gyr). Conservatively, we consider the effective annihilation volume $V$ to be the volume of the celestial body BD/NS. For NS, and for both $s-$ and $p-$ wave DM, $t_{\rm eq}$ will be smaller than 1 Gyr for scattering cross sections of $\mathcal{O}$(10$^{-50}$ cm$^2$) and higher, which is much lower than the sensitivity for NS as shown in Fig. \ref{fig:limits}. For BDs, the volume within which annihilation takes place is larger, because BDs have larger radii than NSs. As such, $t_{\rm eq}$ for BDs is generically longer. For $s-$wave DM, equilibrium can be reached within $\mathcal{O}$(1 Gyr) if the scattering cross section is greater than $\mathcal{O}$(10$^{-41}$ cm$^2$) for all DM masses up to 10$^2$ TeV, which is approximately the testable cross section with BDs in Fig. \ref{fig:limits}. For $p$-wave, DM masses up to 10 GeV can reach equilibrium for scattering cross sections of 10$^{-40}$ cm$^2$ and higher, while 10$^5$ GeV can equilibrate only for cross sections higher than 10$^{-37}$ cm$^2$. However, we emphasize again that these equilibriation timescales will be much faster if the DM thermalizes within the celestial body and settles into a thermal volume. Therefore, except for very high mass (more than 10$^5$ GeV) $p-$wave DM in BDs, the equilibration condition is justified for the parameter space covered in Fig. \ref{fig:limits}.

\subsection{Galactic Center Excess}
\label{subsec:gce}

Our results are particularly interesting in light of observations of a $\gamma$-ray excess of unknown origin emanating from the Galactic center, the ``Galactic Center Excess". The origin of this excess is not yet known~\cite{Goodenough:2009gk,Daylan:2014rsa,Leane:2019xiy,Zhong:2019ycb,Leane:2020nmi,Leane:2020pfc,Buschmann:2020adf,Abazajian:2020tww,List:2020mzd,DiMauro:2021raz}. We note that the BD-focused annihilation signal could potentially explain the GCE, as the 100 GeV DM signal is within the normalization of the GCE (and the annihilation of ~100 GeV DM can produce the correct gamma-ray spectrum). 

Such a signal would be particularly interesting as it provides a density scaling that does not follow annihilating DM, but instead follows the DM density multiplied by the local BD number density. While the morphology of the galactic center excess is not definitively known, some recent work claims that it may be consistent with the morphology of the stellar bulge~\cite{Abazajian:2020tww} (though see e.g. Ref.~\cite{DiMauro:2021raz}), a result which has been used to conclude that astrophysics, rather than DM, powers the excess. Our setup, on the other hand, could potentially explain such a morphology with a DM origin. 

For this combination of mass and cross section parameters, our results would also predict a bright $\gamma$-ray signal from the Sun, which is not observed~\cite{Leane:2017vag,Albert:2018jwh}. However, these constraints can be broadly evaded, depending on the particle decay lengths. For example, the solar limits can be decreased if the mediator has a decay length that is shorter, and is therefore extinguished in the Sun (which is larger) but escapes the BD. More generally, the solar limits can be evaded with a decay length that is much longer than an AU, as this suppresses the flux from the Sun and depending on mediator boost can also enlarge the signal's angular region. In any case, it is also important to note that the direct detection limits overlap with the favored GCE parameter space (and are stronger by about an order of magnitude), so this GCE explanation would only be valid for classes of models with slightly suppressed DD rates, and non-suppressed annihilation rates.

Finally, the NS-focused annihilation signal cannot be responsible for the Galactic Center GeV gamma-ray Excess or the overlapping anti-proton excess~\cite{Cuoco:2017rxb,Cholis:2019ejx,Cuoco:2019kuu,Hooper:2019xss}. This is simply because the flux produced by NSs is low, and both the GCE and anti-proton excess have substantially larger rates.

\section{Signals in Globular Clusters}
\label{sec:globular}

Globular clusters are very dense stellar systems. They have the typical mass of dwarf galaxies, but their size is a $\mathcal{O}(10)$ factor smaller. They can be found in the halo or bulge regions of galaxies. Neutron stars can exist in the center of globular clusters, while BDs may be mostly expelled into the halo~\cite{Griest:1989wd,2019:Caiazzo,2019:Dieball} because of mass-segregation. As such, in this section we study the prospects of `NS-focused' annihilation in globular clusters. We will focus on the globular cluster Tucanae 47 (also called ``Tuc 47"), as it is relatively close by, massive (and so contains a high number of NSs), and is expected to be DM dense. While we expect the globular cluster signal to be weaker than that from the Galactic center, it may provide a corroborating signal in case a detection is first made in the Galactic center. We also expect that new clusters will be found, which may improve the sensitivities compared to Tuc 47 observations.

\subsection{Dark Matter in Globular Clusters}

In order to study the DM capture from NS in globular clusters, we first need to calculate the DM density in globular clusters. Although it is currently impossible to do this with any great accuracy, developments in observation and simulated evolution of globular clusters embedded in Galactic halos can provide an estimate of the DM content. Some time ago, in Ref.~\cite{1984:Peebles}, it was suggested that globular clusters might be formed in subhalos of DM before falling into Galactic halos. Observations of $O(1)$ mass-to-light ratios and tidal stripping from stars from some globular clusters suggest that a significant DM component cannot reside with the observed stellar distribution~\cite{Moore:1996}. These observations set an upper limit on the DM content of globular clusters. 

Simulations have suggested how the above observations can be reconciled with a scenario of globular cluster formation via tidal stripping. Ref.~\cite{Gao:2004} suggested a scenario where continuous mass-loss occurs via tidal stripping once a subhalo falls into a larger halo. In this process, the orbit of the subhalo decays down towards the centre of the larger halo. The tidal stripping of DM from old globular clusters has been studied with $N$-body simulations~\cite{Mashchenko:2005,Moore:2006,Creasey:2018bgv,Saitoh:2006,Griffen:2009vg}. These results render support for the scenario of globular cluster formation within DM subhalos that are tidally stripped by the host galaxy. In this manner, it also explains how these globular clusters grow with baryon-dominated cores. 

DM in the core of such globular clusters might have survived tidal stripping until the present time. This assumption is supported by the results of Ref.~\cite{Mashchenko:2005}, where it is seen that the presence of the stellar core makes the subhalos more resilient to tidal stripping. For NFW halos, the innermost DM density is not modified by the external tidal field. Motivated by these results, DM signals from globular clusters has been studied in several works~\cite{Bertone:2007ae,McCullough:2010ai,Zaharijas:2007ci,Amaro-Seoane:2015,Wood:2008,Fortes:2019gfe,Dasgupta:2019}. 

\subsection{Mass and Velocity Distributions for Tucanae 47}

Here, we utilize the well-studied globular cluster Tuc 47 as a template globular cluster for our calculations, noting that future observations may find stronger constraints for alternative systems. The baryonic properties of Tuc~47 are well studied. It has a baryonic mass of $\sim 10^6 M_{\odot}$, a core radius of $r_c$ = 0.5 pc, a tidal radius $r_t$ = 70 pc and a half-light radius, $r_h$ = 3.7 pc~\cite{Giersz:2010nf}. 

For the mass of the DM halo in Tuc 47 cluster, we can use the relation between the current baryonic mass of the globular cluster, and the mass of the initial DM subhalo, $M_{GC} = 0.0038 M_{DM,0}$~\cite{Griffen:2009vg}. This imples that the initial mass of the DM subhalo of Tuc 47 is $\sim 2 \times 10^8 M_{\odot}$. 

To estimate the DM density in Tuc 47, we follow a similar approach to Refs.~\cite{Bertone:2007ae,McCullough:2010ai,Amaro-Seoane:2015}. The original DM halo of Tuc 47 can be modeled using an NFW profile, as per Eq.~\ref{eq:DM_density}. Further inclusion of baryonic feedback leads to an adiabatic contraction of the DM halo \cite{1986:Blumen,Gnedin:2003rj}. However, the DM cusp created by the adiabatic contraction can be shallowed by the heating of DM due to collision with stars~\cite{Merritt:2003qk}, creating a core of constant density~\cite{Merritt:2003qk,Bertone:2007ae,McCullough:2010ai,Amaro-Seoane:2015}. 
The size of this core can be estimated as the radius at which the two-body relaxation timescale becomes greater than the age of the cluster. In Tuc 47, this radius is about $\sim$ 4 pc \cite{Giersz:2010nf}. Following Refs.~\cite{Bertone:2007ae,McCullough:2010ai,Amaro-Seoane:2015}, we fix the DM density in the inner parsecs (within $\sim$ 4 pc) of Tuc 47 to be $\rho_{DM} = 10^3$ GeV/cm$^3$. 

If Tuc 47 hosts a central IMBH \cite{Kltan:2017}, there may be a spike in the DM density in the central regions of the globular cluster \cite{Brown:2018pwq}. This effect is most pronounced for radii less than $0.01$ pc. However, according to recent simulations, most NS will be present only beyond $\sim$ 0.1 pc for \cite{Ye:2019luh}, so even using models with a pronounced DM spike in the inner regions will not change the NS-focused rates significantly.

We assume a velocity dispersion of 10 km/s \cite{Giersz:2010nf}. This is consistent with the fact that globular clusters in general have low stellar velocity dispersions of $v \sim 10-20$ km/s \cite{Baungardt:2018}.

\subsection{Neutron Stars in Globular Clusters}

Globular clusters are known to be efficient at producing millisecond pulsars (MSPs). Multiple surveys have found 157 pulsars in 30 globular clusters, including 38 in Terzan 5 and 25 in Tuc 47 (see e.g. Refs.~\cite{2005:Camilo,2008:ransom}). Although globular clusters make up only about 0.05\% of stars in the Milky Way, collectively, globular clusters contain more than one third of known MSPs in our Galaxy \cite{Manchester:2004bp}. Globular clusters also contain many low-mass X-ray binaries (LMXBs) with NS acceretors \cite{1975:Clark}. 
 
 The large number of MSPs and LMXBs suggest that a typical Galactic globular cluster on average contain at least a few 100 NSs. The high numbers of NSs seen in globular clusters is in tension with the fact that NSs may be born with large natal kicks~\cite{2005:Hobbs} when the formation occurs from core-collapse supernovae (CCSNe). However, the discovery of the high-mass X-ray binaries with long orbital periods and low eccentricities \cite{Pfahl:2001df} suggests that some NSs must be born with very small natal kicks. More recent studies have suggested that electron-capture supernovae (ECSNe) can solve the retention of NS problem by producing many NS with small kicks \cite{Ivanova:2008,2004:Pods}. These studies showed that a large number of NSs could be retained in globular clusters by formation through ECSNes \cite{Kuranov:2006,2004:Pods,Ivanova:2008}. Simulations \cite{Ivanova:2008,Kremer:2019iul,Ye:2019luh} indicate that $O(100)$ NSs could be retained in a typical globular cluster with mass $\sim$ 10$^5$ M$_{\odot}$. 
 
 We note that for a massive cluster like Tuc-47, the number of NSs that could be retained within the cluster is debated. The range in the number of NSs that could be retained within a Tuc-47 like cluster cited in literature lies between $\sim$ $300-4000$ \cite{Heinke:2005,Ivanova:2008,2004:Pod,Ye:2019luh}. We assume that there are $\sim$ 10$^3$ NSs retained in Tuc-47 motivated by recent numerical simulations that takes into account ECSNe formation of NSs with small natal kicks in a Tuc-47 like cluster (model 26 of \cite{Ye:2019luh}). However, while interpreting our results, this uncertainty in NS numbers should be kept in mind.
 
\subsection{Neutron-Star Focused vs. Standard Halo Annihilation}

\begin{figure} [t!]
    \centering
    \includegraphics[width=\columnwidth]{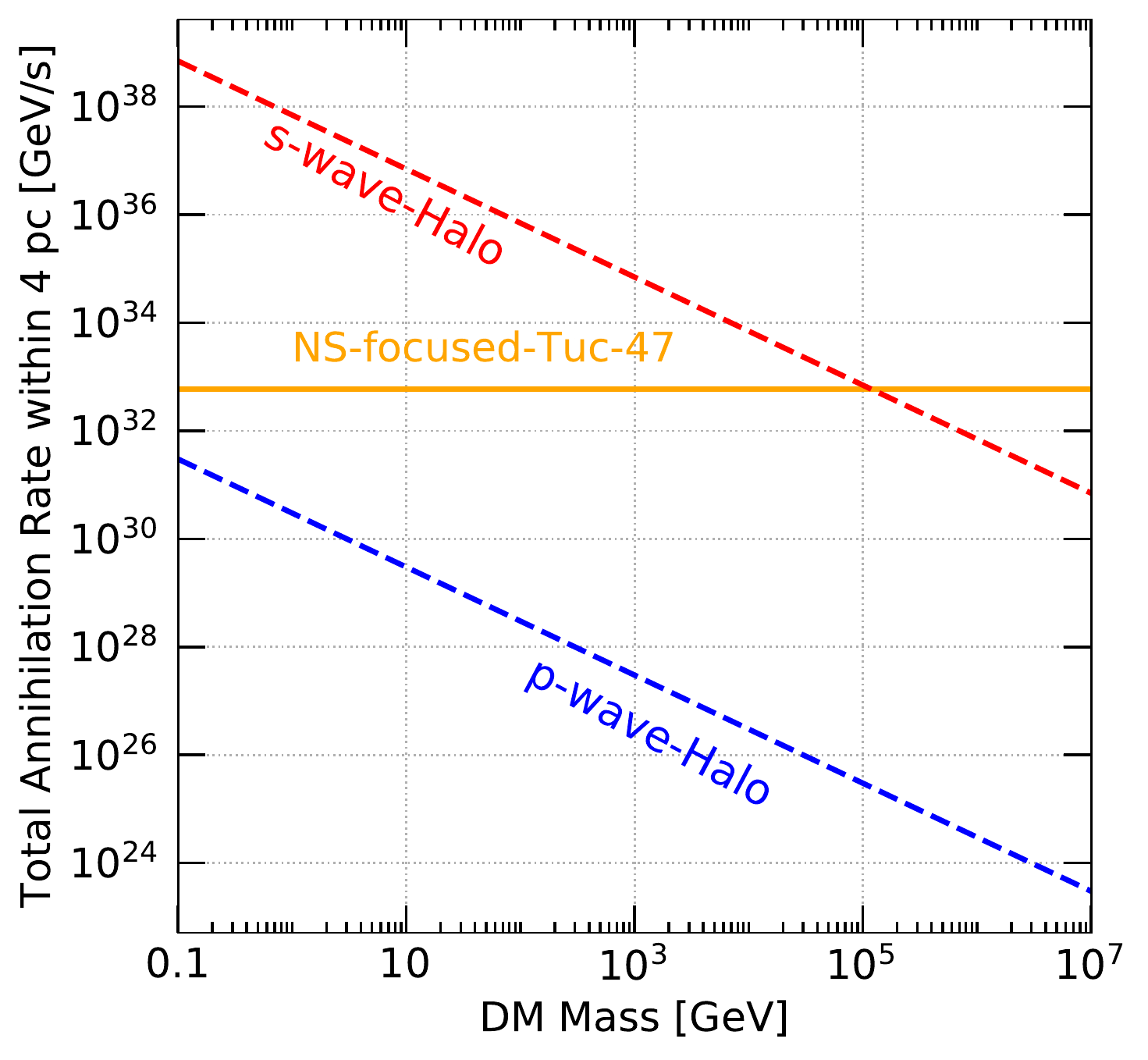}
    \caption{NS-focused annihilation vs. halo annihilation signals in globular cluster Tuc 47, for $s-$wave and $p-$wave DM, for varied DM masses.}
    \label{sensitivity_cluster}
\end{figure}

 We now calculate our NS-focused signal for Tuc 47. We integrate over the inner 4 pc of Tuc 47 using an assumed constant DM density of $10^3$ GeV/cm$^3$, and $\sim 10^3$ NSs. The 4 pc integration boundary is chosen because we assume that most of the NSs are confined within this region of Tuc-47, roughly consistent with the results of Ref.~\cite{Ye:2019luh}. Similar to the total NS numbers, the NS radial number distributions in the central regions of globular clusters is not well known and is a topic of active research. We emphasize that for this globular cluster signal, we are taking a number of well-motivated estimates (rather than definitively known quantities), to simply demonstrate how the population of NSs in globular clusters may provide NS-focused annihilation signals. 
 
 Figure~\ref{sensitivity_cluster} shows the relative strength of the NS-focused annihilation rate for Tuc 47, compared to $s$- and $p$-wave halo annihilation within the integration volume. We see that the NS-focused annihilation can be orders of magnitude higher than the $p-$wave standard halo annihilation signal. In the globular cluster case, $p-$wave annihilations are even more suppressed due to the especially low velocity dispersions in these systems. In fact, if the DM is sufficiently massive, greater than $10^5$ GeV, then the `NS-focused' signal can even dominate over the standard $s-$wave annihilation signal within the integration volume. 
 
  \textit{Fermi} has seen bright $\gamma$-ray emission from Tuc 47 \cite{Abdo:2009rjc}. This emission generally has been attributed to millisecond pulsar emission, though there is some debate whether DM can contribute to it \cite{Brown:2018pwq,Bartels:2018qgr,Brown:2019teh}. The `NS-focused' signal described in Fig. \ref{sensitivity_cluster} is too low to explain this emission from Tuc 47. In general, we note that the `NS-focused' signal from Tuc 47 is below the reach of present \textit{Fermi}-LAT sensitivity \cite{FERMI:2020}. In the future, a more optimal signal from a different globular cluster may be found.
   
\section{Summary and Conclusions}
\label{sec:conclusion}

Indirect DM searches have typically focused on either the effects of DM-DM interactions (e.g., annihilation) or DM-SM interactions (e.g., DM scattering off compact objects). In this work, we have demonstrated new detection possibilities which combine these interactions, and allow celestial body populations to ``focus" DM-DM interactions and significantly enhance their rate. 

The signals from such processes have several new phenomenological features. The signal (i) distinctively scales linearly with DM density, rather than with the squared DM density like standard halo annihilation, and (ii) can dominate over standard halo annihilation in some situations, especially when the annihilation rate is velocity suppressed. This is particularly valuable in the case of suppressed halo annihilation (e.g. $p$-wave annihilation), which would otherwise be undetectable. This signal requires that DM annihilates to a sufficiently long-lived mediator, in order to allow products to escape the celestial object and be detectable with indirect detection experiments.

We surveyed the potential celestial object targets for this signal, and identified neutron stars and brown dwarfs as ideal targets. Neutron stars can be a particularly interesting laboratory, due to their extreme densities and gravitational wells. A standard scenario considered in the literature is DM that scatters and is captured by neutron stars, and subsequently either heats the neutron star or collapses the entire system into a black hole. In contrast, for the first time, we have considered the indirect detection signals that arise from neutron stars. Similarly, brown dwarfs are quite dense, and produce larger $\gamma$-ray fluxes (compared to neutron stars) for relatively-large DM scattering cross sections, due to their large radii and larger number density in our Galaxy. For the first time, we considered DM annihilation to long-lived particles within brown dwarfs.

We studied this focusing signal in two different environments; the Galactic center, and in the globular cluster Tuc 47. For the Galactic center, we found that for NSs, the focused rate can exceed the p-wave rate for masses $m_{\chi} > 10^3$ GeV. For brown dwarfs, we found the focused rate can exceed the p-wave suppressed flux from halo annihilation for $m_{\chi} > 0.1$ GeV and even s-wave annihilation for $m_{\chi} \gtrsim 10^3$ GeV . 

We pointed out that brown-dwarf-focused annihilation may also explain the Galactic Center Gamma-Ray Excess, and can do so with a morphology partially scaling with the stellar matter, providing a DM interpretation of the excess using typically non-DM morphologies. 

When studying globular cluster Tuc 47, we pointed out that while the Galactic center signal could be stronger, the globular cluster signal could be used as a corroborating check in the case a signal is first seen in the Galactic center. We also pointed out that, importantly, new globular clusters may be found, and even better sensitivities will be possible using this framework for more optimal clusters.

For the first time, we have set limits on DM annihilation to long-lived particles in brown dwarfs and neutron stars, and the resulting DM scattering cross sections. In different parts of parameter space, we outperform both existing solar limits, and direct detection experiments. Using \textit{Fermi}'s measured Galactic center gamma-ray fluxes, brown dwarfs provide the strongest new limits, with an improvement in the sub-GeV mass range up to nine orders of magnitude. This is due to comparably poor direct detection sensitivity in the sub-GeV mass regime, and the cooler cores of brown dwarfs which allow DM to not evaporate, unlike the Sun. We showed that depending on model assumptions, neutron stars can potentially outperform spin-dependent direct detection by $\sim4-8$ magnitude, in the TeV DM-mass range. In this mass range, neutron stars can potentially even outperform the Sun, by $\sim1-2$ orders of magnitude.

\section*{Acknowledgments}
We thank Nicole Bell, Alexander Friedland, Carolyn Kierans, Shirley Li, Smadar Naoz, and Juri Smirnov for helpful discussions. RKL, PM and NT are supported in part by the U.S. Department of Energy under Contract DE-AC02-76SF00515. TL is partially supported by the Swedish Research Council under contract 2019-05135, the Swedish National Space Agency under contract 117/19 and the European Research Council under grant 742104.

\bibliography{bibliography1}

\end{document}